\documentclass[aps,prx,twocolumn,superscriptaddress,groupedaddress]{revtex4-2}

\usepackage{graphicx}
\usepackage{dcolumn}
\usepackage{bm}
\usepackage{mathtools}
\usepackage[colorlinks=true]{hyperref}
\usepackage[dvipsnames]{xcolor}
\usepackage[mathlines]{lineno}

\newcommand\myshade{85}
\colorlet{mylinkcolor}{Red}
\colorlet{mycitecolor}{Blue}
\colorlet{myurlcolor}{Gray}

\hypersetup{
  linkcolor  = mylinkcolor!\myshade!black,
  citecolor  = mycitecolor!\myshade!black,
  urlcolor   = myurlcolor!\myshade!black,
  colorlinks = true
}

\newlength{\thelinewidth}
\thelinewidth=\textwidth

\begin{document}

\title{Search of spin-dependent fifth forces with precision magnetometry}

\author{N.~Crescini}\altaffiliation[Present address: ]{IBM Research-Zürich, Säumerstrasse 4, CH-8803 Rüschlikon, Switzerland}
	\email{ncr@zurich.ibm.com}
	\affiliation{INFN-Laboratori Nazionali di Legnaro, Viale dell'Universit\`a 2, 35020 Legnaro (PD), Italy}
	\affiliation{DFA - Università degli Studi di Padova, Via Marzolo 8, 35131 Padova, Italy}

\author{G.~Carugno}
	\affiliation{DFA - Università degli Studi di Padova, Via Marzolo 8, 35131 Padova, Italy}
	\affiliation{INFN-Sezione di Padova, Via Marzolo 8, 35131 Padova, Italy}

\author{P. Falferi}
	\affiliation{IFN-CNR, Fondazione Bruno Kessler, and INFN-TIFPA, Via alla Cascata 56, 38123 Povo (TN), Italy}	

\author{A.~Ortolan}
	\affiliation{INFN-Laboratori Nazionali di Legnaro, Viale dell'Universit\`a 2, 35020 Legnaro (PD), Italy}

\author{G.~Ruoso}
	\affiliation{INFN-Laboratori Nazionali di Legnaro, Viale dell'Universit\`a 2, 35020 Legnaro (PD), Italy}

\author{C.C.~Speake}
	\affiliation{School of Physics and Astronomy, University of Birmingham, West Midlands B15 2TT, UK}

\begin{abstract}
Spin-dependent fifth-forces are associated with particles beyond the standard model. In particular, light pseudo-scalar bosons mediate long-range forces, allowing mass to interact with spins.
The search of these interactions can be performed by periodically varying the distance between a source mass and a spin ensemble, in order to modulate the force intensity and detect it with precision magnetometry techniques. In our setup the force arises from room temperature lead masses and is detected in a paramagnetic crystal at 4.2\,K, whose magnetisation is monitored by a SQUID-based magnetometer with the sensitivity of $53\,\mathrm{aT/\sqrt{Hz}}$. Our measurement places the most stringent constraints on spin-mass interactions in the ranges 1\,cm to 10\,m and 10\,km to 300\,km, with couplings $g_p^e g_s^N \le 5.7\times 10^{-32}$ and $g_p^e g_s^e \le 1.6\times 10^{-31}$ at 95\% C.L., improving existing limits up to more than two orders of magnitude. We show that this experimental technique may be further leveraged to explore a vast region of the fifth force's parameter space, with an interaction range longer than a few centimetres.
\end{abstract}

\maketitle

\section{Introduction}
\label{sec:intro}
The known fundamental forces of nature are four: electromagnetism, weak interaction, strong interaction, and gravity. The standard model of particle physics is a remarkably successful theory that unifies the first three, and explains the forces as the exchange of the respective gauge bosons, i.\,e. photons, $Z^0$ and $W^\pm$, and gluons \cite{bettini_2008,griffiths}. Grand unification theories, like string theory, also describe the gravitational interaction as mediated by a graviton \cite{Feynman_2003,doi:10.1142/S0217751X96002583}. Strong and weak interactions are called short-range forces, since they act in a typical length below $10^{-15}\,\mathrm{m}=1\,\mathrm{fm}$, the size of a nucleus, while electromagnetic and gravitational interactions have infinite range. The mass of the mediator $m$ determines the force range through the Compton wavelength
\begin{equation}
\lambda=\frac{h}{mc},
\label{eq:compton}
\end{equation}
where $h$ is the Planck constant and $c$ is the speed of light in vacuum. The $Z^0$ and $W^\pm$ bosons are heavy, so the weak interaction strength rapidly decreases approximately above 0.001\,fm. All the other gauge bosons are massless and mediate long-range forces, with the peculiarity of the strong interaction, being effectively short-ranged because of color confinement. Gravity and electromagnetism do not allow for the coupling between spin and mass in the absence of relative motion, implying that a long-range spin-mass force is necessarily a fifth force, not encompassed by the standard model.

\begin{figure*}
\centering
\includegraphics[width=\textwidth]{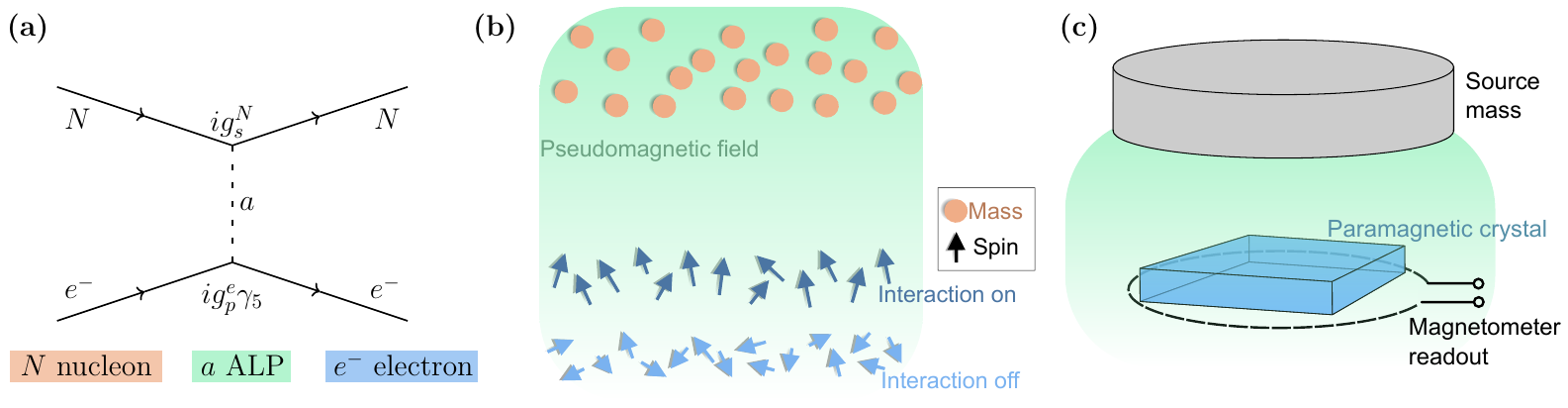}
\caption{Feynman diagram, particle description, and schematic drawing of the experiment. (a) The diagram represents the interaction between a nucleus $N$ and the spin of an electron $e^-$, mediated by an ALP $a$ whose scalar nucleon coupling is $g_s^N$ and pseudoscalar electron one is $g_p^e$. (b) The pink circles are the nuclear masses generating the pseudomagnetic field attracting the spins, which are dark blue (light blue) arrows for $r\ll\lambda$ ($r\gg\lambda$). (c) A lead block (grey) hosts the monopole masses, and the paramagnetic crystal (blue) is the spin detector. The drawing is not to scale, see text for further details.}
\label{fig:1}
\end{figure*}

The existence of ultralight scalar particles is conjectured by a number of theories beyond the standard model, usually as a consequence of the Nambu-Goldstone theorem \cite{weinberg}.
Among these particles the axion is particularly well-motivated, since it arises from the Peccei and Quinn solution of the strong CP problem \cite{pq,weinberg1978new}, and is of cosmological interest as an acknowledged Dark Matter candidate \cite{MARSH20161}. Axions, together with axion-like particles (ALPs), are the subject of a wide and flourishing research, which in the last years encountered a conspicuous expansion \cite{Jaeckel:2010ni,IRASTORZA201889}. 
Along with particle physics, the search of ALP mediated forces is strongly motivated by astrophysics, as at long distances they may, for example, affect structure formation or even overwhelm gravity \cite{MARSH20161,IRASTORZA201889}.
The search of new macroscopic forces as a way to detect ALPs was first suggested by Moody and Wilczek \cite{wilczek}, and triggered a number of experimental efforts \cite{Vorobyov1988146,PhysRevLett.67.1735,PhysRevLett.68.135,PhysRevLett.70.701,PhysRevLett.77.2170,PhysRevLett.82.2439,PhysRevLett.98.081101,PhysRevD.78.092006,PhysRevLett.106.041801,PhysRevLett.115.201801,CRESCINI2017677,Rong2018,PhysRevLett.120.161801}. These are framed in the so called ``pure laboratory'' searches, as they do not rely on any source of force outside the experiment itself, and consist of low-energy precision measurements assessing anomalies of present theories. 
In the plethora of exotic interactions, three different potentials arise from the exchange of a spin-0 boson \cite{PhysRevA.99.022113,PhysRevD.89.114022}: dipole-dipole, monopole-dipole, and monopole-monopole.
Hereafter we focus on the description of monopole-dipole interactions, which macroscopically can be viewed as spin-mass forces; for the case of an electron ($e^-$) nucleon ($N$) interaction an example Feynman diagram is shown in Fig.\,\ref{fig:1}a. 

Despite seemingly compelling arguments for exotic phenomena  \cite{Brumfiel2004}, presently, there is no evidence of a new force of nature. The most stringent limits are essentially obtained with torsion balances \cite{PhysRevLett.70.701,PhysRevLett.77.2170,PhysRevLett.98.081101,PhysRevD.78.092006,PhysRevLett.106.041801,PhysRevLett.115.201801}, comagnetometers \cite{PhysRevLett.68.135,PhysRevLett.120.161801} and magnetic materials coupled to SQUIDs \cite{Vorobyov1988146,PhysRevLett.82.2439,CRESCINI2017677}, or likewise with other experimental techniques which have been proposed \cite{PhysRevD.91.102006,PhysRevLett.113.161801,chen2019ultrasensitive} or already applied \cite{PhysRevLett.67.1735,Rong2018}.
Moreover, the analysis of combined pure laboratory and astrophysical limits \cite{PhysRevD.86.015001}, and the study of atomic and molecular electric dipole moments \cite{PhysRevLett.120.013202} gave outstanding results in toughening the bounds on monopole-dipole interactions.
Pure laboratory searches of fifth forces remain of absolute importance for the understanding of fundamental physics, and, with this work, we indicate a scheme which drastically improves present experimental constraints, and has the potential to overcome astrophysical bounds and test present axion models.

The use of a magnetic material as a detector of spin-dependent forces was pioneered by Vorobyov and Gitarts \cite{Vorobyov1988146}, followed by Ni and collaborators \cite{PhysRevLett.82.2439}, who held the best limits on this type of interaction for more than a decade. More recently, a measurement by our group \cite{CRESCINI2017677} improved these constraints with a pilot setup, which demonstrated the effectiveness of our experimental configuration. In this work we further develop the apparatus, obtain much improved results, and hence demonstrate the extraordinary sensitivity that can be attained by searching for fifth forces with precision magnetometry. Below, we show how a larger size apparatus may evade the main fundamental noises of the scheme, allowing the exploration of a much wider range of the monopole-dipole parameter space.

\begin{figure*}
\centering
\includegraphics[width=\textwidth]{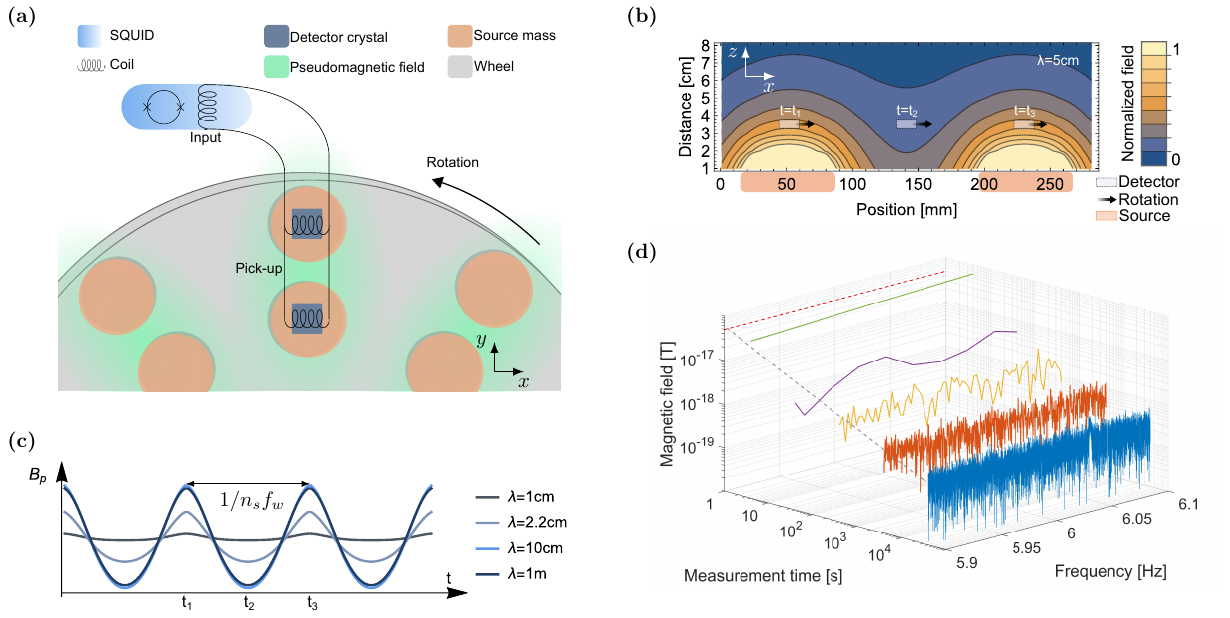}
\caption{Signal modelling and integrated magnetic background. (a) Scheme of the experimental apparatus (not to scale). The paramagnetic crystals (dark blue), the pick-up superconducting coils, and the SQUID are at liquid helium temperature. The room temperature part of the setup are the lead masses (orange) fixed to the wheel (grey), which, when rotating, modulate the source-detector distance (see Appendix \ref{app:a} sections for more details). 
(b) Contour plot of the pseudomagnetic field generated by two neighbouring source masses, where the rectangles show the detector and source positioning. At the times $t_1$ and $t_3$ the signal is maximum, while at $t_2$ it is minimum. 
(c) Time dependence of the pseudomagnetic field for different Compton wavelengths. Different tones of blue show the signal modulation due to some example wavelengths, which are calculated with the numerical integration of Eq.(\ref{eq:bint}). The times $t_{1,2,3}$ correspond to those of the previous figure.
(d) Solid lines are the magnetic field spectra in the frequency band of the signal for integration times ranging from 4\,s (green) to 11\,h (blue). The grey dashed line shows the $1/\sqrt{t}$ trend of the noise, which is respected throughout the entire measurement, and the red dashed line is the SQUID background estimation.}
\label{fig:2}
\end{figure*}

\section{Fifth force signal}
\label{sec:signal}
The non-relativistic monopole-dipole interaction related to the Yukawa couplings in the diagram of Fig.\ref{fig:1}a reads
\begin{equation}
V(\mathbf{r}) = \frac{\hbar^2 g_p^e g_s^N}{8\pi m_e} \hat{\boldsymbol{\sigma}}\cdot \hat{\mathbf{r}} \Big( \frac{1}{\lambda r} + \frac{1}{r^2} \Big)e^{-r/\lambda},
\label{eq:v}
\end{equation}
where $\hbar=h/2\pi$, $ g_p^e$ and $g_s^N$ are the pseudoscalar and scalar couplings of the ALP to electrons and nucleons, $m_e$ is the electron mass and $\hat{\boldsymbol{\sigma}}$ is its Pauli vector, and $\mathbf{r}=r\,\hat{\mathbf{r}}$ is the spatial vector connecting the nucleon to the spin. The potential violates parity and time-reversal and hence it is not CP conserving.
Eq.\,(\ref{eq:v}) states that for $r<\lambda$ and positive $g_p^e g_s^N$, the minimum energy configuration is the spin pointing towards the mass, condition which is exponentially relaxed when $r>\lambda$, as shown in Fig.\,\ref{fig:1}b. In this sense, using the Bohr magneton $\mu_B$, the potential may be recast as a pseudomagnetic field acting on an electron spin $V(\mathbf{r})=-\mu_B \hat{\boldsymbol{\sigma}}\cdot\mathbf{b}(\mathbf{r})$, yielding
\begin{equation}
\mathbf{b}(\mathbf{r}) = -\frac{\hbar g_p^e g_s^N}{4\pi q} \hat{\mathbf{r}} \Big( \frac{1}{\lambda r} + \frac{1}{r^2} \Big)e^{-r/\lambda},
\label{eq:b}
\end{equation}
where $q$ is the charge of the electron. The field $\mathbf{b}(\mathbf{r})$ is not mediated by photons, so it does not respect Maxwell's equations, and is not screened by superconducting shields, allowing us to place the detector in a magnetically controlled environment without reducing the interaction intensity. Moreover, we can take advantage of its polynomial and exponential dependence on $r$ to modulate the interaction strength, and thus the searched for signal, by varying the spin-mass distance (see Fig.\,\ref{fig:1}b).

A typical apparatus to probe the pseudomagnetic field $\mathbf{b}(\mathbf{r})$ is formed by an ensemble of nuclei in a macroscopic mass, called source, and a collection of spins composing the detector.
To calculate the pseudomagnetic field generated by the source mass, one needs to integrate Eq.\,(\ref{eq:b}) over its volume $\Omega$. Assuming a uniform nuclear density $\varrho_n$, the field at the detector position results
\begin{equation}
B_p = \varrho_n \int_\Omega \mathbf{b}(\mathbf{r}) \mathrm{d}\Omega,
\label{eq:bint}
\end{equation}
whose integration can be performed by numerical or analytical means \cite{PhysRevLett.82.2439,CRESCINI2017677,Rong2018}.
Microscopically, the field $B_p$ effectively tilts the spins in the direction of the mass, so macroscopically it changes the detector's magnetisation \cite{landau}, which can be detected with a magnetometer, as displayed in Fig.\,\ref{fig:1}c.
In this sense the magnetic material turns a pseudomagnetic field into a real one through its magnetic susceptibility $\chi$, and the arising signal is
\begin{equation}
\mu_0 M=\chi B_p,
\label{eq:m}
\end{equation}
which, for simplicity, is expressed in units of Tesla using the magnetic permeability of vacuum $\mu_0$.

\section{Experimental setup}
\label{sec:setup}
As the magnetic sensitivity is a key element for this type of searches, we feature in our apparatus a SQUID-based magnetometer, which is among the most sensitive sensors available. 
State-of-the-art SQUIDs are affected by a flux noise of about $1\,\mu\phi_0/\sqrt{\mathrm{Hz}}$, where $\phi_0\simeq 2\times10^{-15}\,$Wb is a flux quanta. It is well known \cite{squidhand} that a SQUID magnetometer can be configured to optimize the field sensitivity over the spatial resolution, as was established by J.E.\,Zimmerman \cite{doi:10.1063/1.1659798,squidhand} and is presently used in different fundamental and applied physics experiments \cite{doi:10.1063/1.4976823,PhysRevLett.122.121802,shaft}. 
This improvement is effective as long as the searched for field is distributed over the whole pick-up coil area. In our case, this means that the magnetometer coils have to be filled with the magnetic sample, while the $B_p$ monopole source must be provided by sizeable masses.

The magnetic samples used as detectors are two paramagnetic $\mathrm{Gd_2 Si O_5}$ (GSO) crystals \cite{gso} with a magnetic susceptibility $\chi\simeq0.7$ \cite{Crescini2017109}, and dimensions $2.5\times 2.5 \times 1.1\,\mathrm{cm}^3$.
The GSO crystals fill two superconducting pick-up coils wired to a SQUID as shown in Fig.\,\ref{fig:2}a. The two inductances are connected in parallel, to match the input impedance of the SQUID and maximise the magnetic field sensitivity of the magnetometer \cite{squidhand}.
The overall system calibration relies on the measurement of the flux-to-voltage conversion coefficient of the SQUID sensor, and on the calculation of the field-to-flux transduction of the pick-up coils. See Fig.\,\ref{fig:2}a for the circuit scheme and Appendix \ref{app:b} for further details.
The source producing the ALPs signal consists of $n_s=12$ pairs of masses mounted on an aluminum disk of about 1\,m diameter.
The total 24 masses are lead cylinders of 6\,cm height and 5\,cm diameter, whose nucleon density is $\varrho_n\simeq 6.8\times 10^{30}\,\mathrm{m}^{-3}$. 
The pairs are evenly spaced over the disk circumference, and each of the masses composing the pair is mounted on the same radius, as depicted in Fig.\,\ref{fig:2}a.
A rotation of the resulting wheel at a frequency $f_w$ varies the source-detector distance with a frequency $n_s f_w$. 
An oscillating $B_p$ periodically orients the spins toward the masses, resulting in a modulation of the GSO's magnetisation detected by the SQUID. The component of the SQUID spectrum at the frequency $n_s f_w$ could thus contain the pseudomagnetic field signal.
The source of this setup is at room temperature, while detector is placed in a controlled environment, namely, a liquid helium cryostat at a temperature of 4.2\,K. 
The cryostat is enclosed in two $\mu$-metal shields, while the detector is surrounded by two additional layers of superconducting NbSn shields with a screening factor of order $10^{10}$. We achieved a minimum source-detector distance of 3.5\,cm between the GSO's centre of mass and the lead cylinders surface.
The pseudomagnetic field is unaffected by the layers screening the electromagnetic radiation, allowing us to model its amplitude ignoring them, as is reported in Fig.\,\ref{fig:2}b.
Numerically solving Eq.\,(\ref{eq:bint}) in the case of our experimental configuration, we obtain the shape and amplitude of the expected signal generated by the wheel rotation, as shown in Fig.\,\ref{fig:2}c.

The wheel rotation is driven by a brushless motor synchronous with a function generator, guaranteeing a stable and persistent signal. We use a capacitive sensor to monitor the sources rotation, and verify that its frequency does not change more than 20 parts per million in more than ten hours of measurement, assuring a coherent integration time which exceeds the run time determined by the cold time of the cryostat.

A significant advantage of this approach lies in the control of the signal frequency, which allows us to operate the setup where the magnetic noise is favourable and, as lower bound, matches the one of the SQUID.
We choose to work at a sweet spot at $n_s f_w\simeq6$\,Hz by rotating the wheel at $f_w\simeq0.5\,$Hz, at this frequency the magnetic field noise of the SQUID-based magnetometer reaches the value of $53\,\mathrm{aT/\sqrt{Hz}}$, as shown in the spectra of Fig.\,\ref{fig:2}d.
Within the uncertainties, this background is compatible with our estimation, primarily consisting of the SQUID noise, and with a smaller contribution likely due to the crystals' magnetisation noise. The reader is referred to Appendix \ref{app:c} section for a detailed noise budget.
We leverage the fact that the pseudomagnetic signal frequency is stable within $1/t$, the measurement duration, to keep narrowing the measurement bandwidth and hence reduce the noise as $\sqrt{t}$. 

A spurious effect which mimics the pseudomagnetic signal is the modulation of a static magnetic field related to the non-zero magnetic susceptibility of lead $\chi_\mathrm{lead}\simeq10^{-5}$; in our setup this AC field is screened. However, exploiting this effect, we performed a consistency test based on standard electromagnetism to further enforce our result. 
A 30\,cm-diameter coil made of a copper wire is used to produce a static field. It is placed below the rotating wheel and parallel to it, to immerse the source masses in a field of $\simeq1.3\,$mT. Without shielding this would generate an AC field of order 10\,nT on the GSO. 
At the SQUID input, we resolved a signal at the expected frequency $f_w n_s$ with an amplitude of 3.3\,aT, compatible with the shielding factor of our setup. 
The magnetic field of the Earth measured at the lead masses is more than 50 times lower than the applied one, assuring that its systematic contribution in our measurement will be below 0.07\,aT.
See Appendix \ref{app:d} for further details.

\section{Fifth force search}

\begin{figure*}
\centering
\includegraphics[width=.9\textwidth]{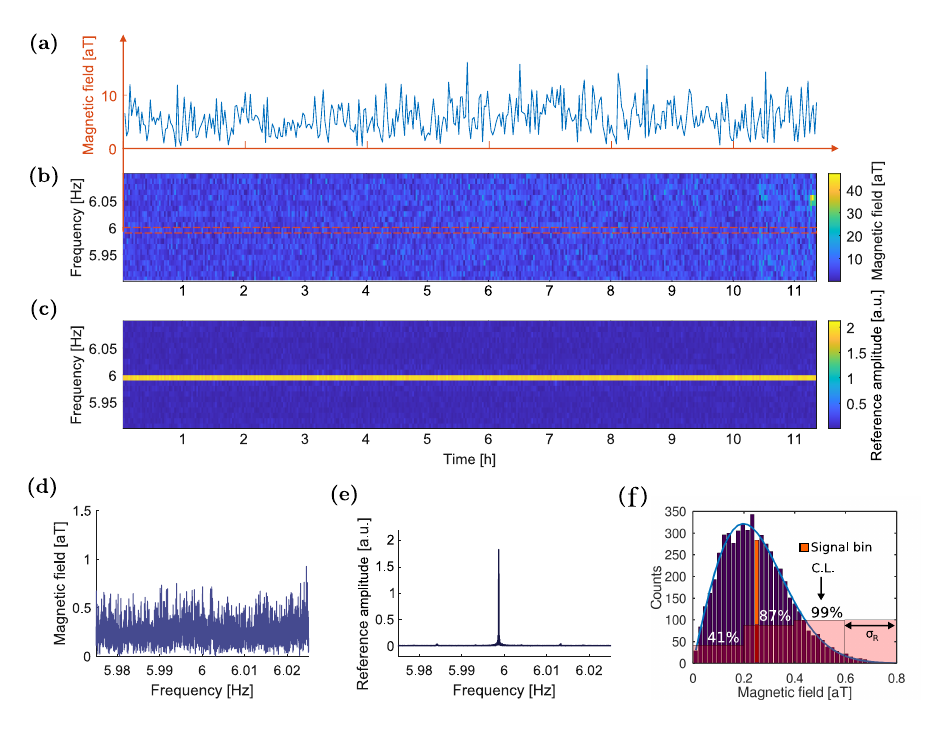}
\caption{
Result of the 11 hours run to search for pseudomagnetic forces. Panels (b)  and (c) show the spectrograms obtained dividing the acquired data in 72 seconds blocks: zoomed frequency regions around the frequency of interest are shown.  Panel (c) shows the reference signal contained in a single bin all along the measurement time. The same bin is delimited with red lines in the SQUID signal spectrogram of panel (b), whose amplitude versus time is shown in panel (a). A single FFT over the entire run for the magnetic field and reference amplitudes  in the frequency region of interest  are shown in panels (d) and (e) respectively. Panel (f) shows a histogram of the corresponding noise spectral amplitudes fitted to a Rayleigh distribution, with indication of the bin containing the value measured at the frequency of the searched for signal. The boxes at the bottom indicates the cumulative probability for multiples of the scale parameter. See text for details. }
\label{fig:3}
\end{figure*}

\begin{figure}
\centering
\includegraphics[width=.5\textwidth]{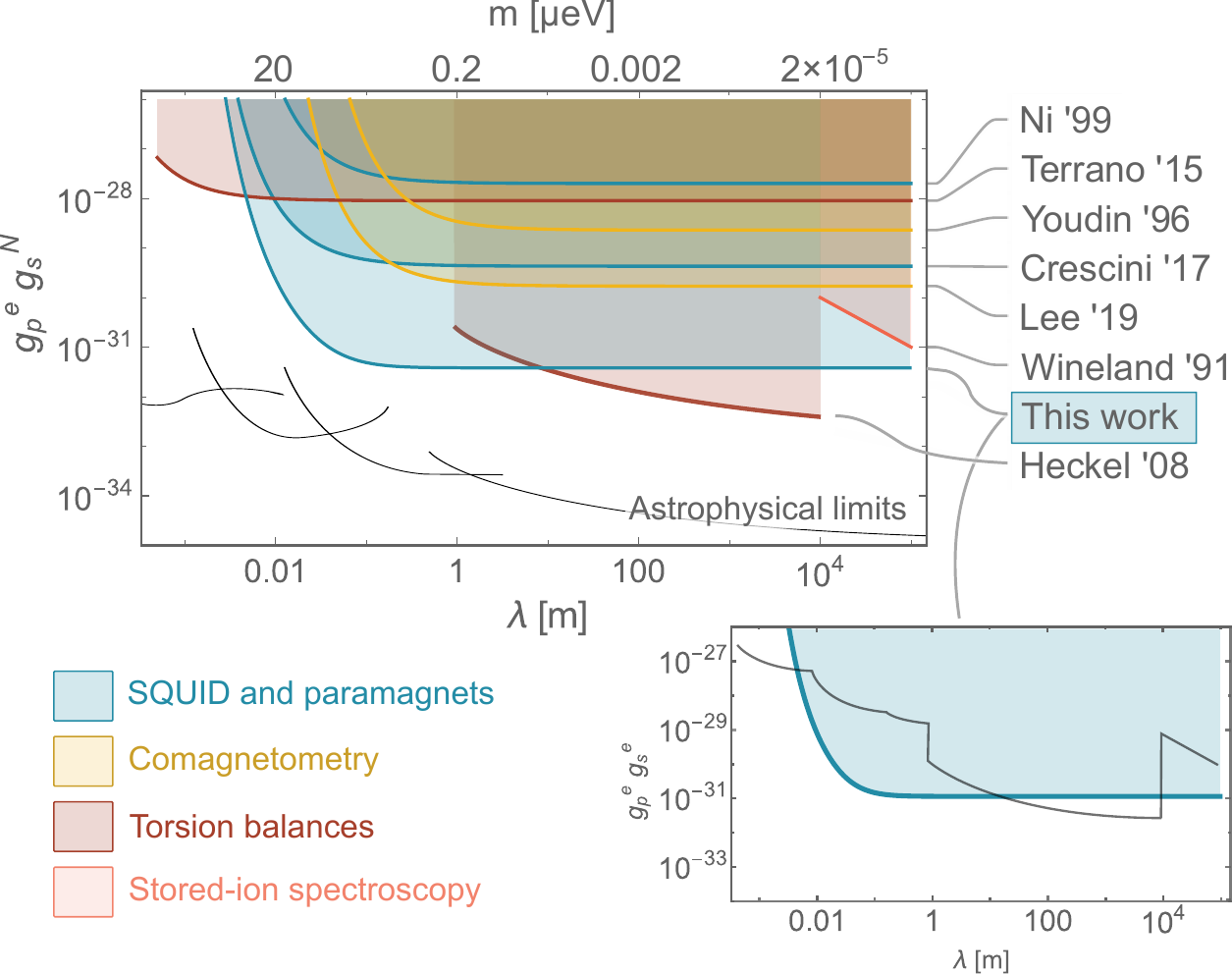}
\caption{Limits on spin mass interaction constants $g_p^e g_s^N$ (upper) and $g_p^e g_s^e$ (lower), derived from the present measurement and compared to the literature \cite{Vorobyov1988146,PhysRevLett.67.1735,PhysRevLett.68.135,PhysRevLett.70.701,PhysRevLett.77.2170,PhysRevLett.82.2439,PhysRevLett.98.081101,PhysRevD.78.092006,PhysRevLett.106.041801,PhysRevLett.115.201801,CRESCINI2017677,Rong2018,PhysRevLett.120.161801}. In the upper plot, the solid lines mark the experimentally excluded areas of the parameter space, and their color depends on the technique used for the measurement. The black lines are mixed laboratory and astrophysical constraints calculated by G. Raffelt \cite{PhysRevD.86.015001}. In the lower plot, a solid blue line represent the result of this work, while the black line are the previous limits, obtained scaling the ones of the electron-nucleon interaction under the assumption that $A=2.5 Z$ for all the source masses.}
\label{fig:4}
\end{figure}

A magnetic measurement was performed while the source wheel of the setup was rotating: this can be directly translated into a measurement of the oscillating pseudomagnetic field generated by the lead masses.  In a first analysis, FFTs of the acquired data are calculated every 72\,s, and then collected into spectrograms:   Fig.\,\ref{fig:3}(b) and (c) shows the spectrograms for the SQUID signal and the reference signal in a zoomed region around 6 Hz. It is clear from panel (c) that the turntable rotation is extremely stable, with the reference peak occupying always the same bin along the entire run. The corresponding bin in the SQUID signal has been delimited with red lines in panel (b) and its amplitude is shown in panel (a).  
From the plots in Fig.s\,\ref{fig:3}(a), \ref{fig:3}(b) and \ref{fig:3}(c) one can see that the magnetic spectra approach a white and constant noise in a frequency band close to $n_s f_w$, and is entirely compatible with stochastic fluctuations at the pseudomagnetic field frequency.
This stimulated us to perform a single FFT of the entire run: the magnetic field and reference amplitudes  in the frequency region of interest  are shown in Fig.\,\ref{fig:3}(d) and (e): indeed the reference is still contained in a single bin at $n_s f_w = 5.99$\,Hz. The frequency offset with respect to Fig. \ref{fig:new} is due to changes in the locking point of the feedback controlling the turntable motion. Such offset has no effects in the signal stability and thus also the searched for magnetic signal would appear in the single bin at $n_s f_w$. We now turn into panel (d), showing the magnetic signal. 
In order to evaluate the presence of a  fifth force signal, we assumed that in a small region of 0.02 Hz width around $n_s f_w$  the disturbances are of the same type and origin: we plotted the distribution of the amplitudes in this frequency window and fitted them with a Rayleigh distribution. As shown is panel (f) a scale parameter $\sigma_R= 0.20\,$aT is found and since the value of the amplitude  at $n_s f_w$ is $a_B = 0.26\,$aT,  well inside  the noise distribution,  no spin-mass interaction has been detected.
We can consequently  put limits on the corresponding coupling constant by replacing in Eq.\,(\ref{eq:m})  $\mu_0 M$ with $2.36\,\sigma_R + a_B = 0.73$\,aT, in order to have a 95\% confidence level (see Fig,\,\ref{fig:3}f). 
For $\lambda \gtrsim 10\,\mathrm{cm}$ we found $g_p^e g_s^N \le 5.7\times 10^{-32}$ and $g_p^e g_s^e \le 1.6\times 10^{-31}$, where we have taken into account that both electrons and nucleons are monopolar sources and we have calculated limits for electron-nucleon and electron-electron interactions. The limits for the full range of $\lambda$ are presented in Fig.\,\ref{fig:4}, where they are also compared with previous results on this type of interactions.

As can also be seen by the spectra of Fig.\,\ref{fig:2}(d), a $\sqrt{t}$ decreasing trend of the system noise is observed, up to the total integration time of $4.1\times10^4\,$s. This is consistent with a zero-mean gaussian fluctuation, as expected for a thermal noise. The noise average  extracted from the Rayleigh distribution is $1.25\sigma_R$, showing that in about 11 hours, the minimum persistent field detectable by our magnetometer with a unity signal-to-noise ratio results $0.25\,$aT.
We mention that a phase-dependent regression could improve the sensitivity of the experiment by a factor $\sim\sqrt{2}$, but requires an improved monitoring of the signal phase.

This work places the most stringent upper bounds on spin-mass interactions over orders of magnitude in the force range $\lambda$, or equivalently in the ALP mediator mass $m$. 
Between 1\,cm and 1\,m we advance our previous result \cite{CRESCINI2017677} and the limit obtained with a $^3$He-K comagnetometer \cite{PhysRevLett.120.161801} of roughly two orders of magnitude. Above 1\,m and below 10\,m we improve the result of a torsion pendulum \cite{PhysRevD.78.092006} whose monopole source is the Earth, which exceeds our measurement's sensitivity above 10\,m. For Compton wavelengths of more than 10\,km the latter experiment, due to the possible presence of systematic uncertainties, does not provide constraints, and the limit was set by stored-ion spectroscopy \cite{PhysRevLett.67.1735}.
The present work improves this limit up to range of about 300\,km.
We do not further discuss the ranges $\lambda\le1\,\mathrm{cm}$ and $\lambda\ge 10^3\,\mathrm{km}$, but mention that torsion balances \cite{PhysRevLett.115.201801,PhysRevD.78.092006} provide the best limits.
We note that, although the sensitivity reached in this work lies seven orders of magnitude away from the model-predicted value of the $g_p^e g_s^N$ coupling \cite{IRASTORZA201889}, we believe that further significant improvements to the experimental method can be made.

First we consider direct improvements of the present setup.
The employment of a $LC$ pick-up coil \cite{journal/apl/79/16/10.1063/1.1408276,Vinante:2002nnd} on resonance with the signal would enable us to suppress the SQUID background and to be limited by the magnetisation noise. On the other hand, the resonant circuit needs a quality factor such that its Johnson noise does not exceed the SQUID noise \cite{paolof2}, which at low frequencies becomes a challenging technical problem.
Crystals with a much higher $\chi$ may significantly improve the experimental sensitivity by raising the signal, however, such materials usually have a much higher magnetisation noise.
A simpler solution to enhance the setup is to use a quantum-limited SQUID \cite{journal/apl/79/16/10.1063/1.1408276} to minimize its noise (see Appendix \ref{app:c}), or lowering the working temperature to also reduce the magnetisation fluctuations.
While the latter upgrades basically improve the precision of the magnetometer, we envision an upscaling of the apparatus which simultaneously reduces the SQUID and magnetisation field noises, and increases the expected signal. The simplicity of the apparatus makes it easily scalable, and, as GSO is readily available in large quantities, we can use more than two parallel coils as pick-up and bolster the idea of the present scheme. 
An upgraded apparatus would allow us to use a large source which can increase the signal strength up to two orders of magnitude.
For example, a setup with 50\,cm-diameter source and a detector comprising 200 pick-up loops of 20\,cm diameter, filled with GSO and read by a quantum-limited magnetometer, could in principle have the sensitivity to detect the model-predicted fifth force \cite{IRASTORZA201889}. 
For what concerns the limitations of the scheme, we expect them to be related to technical noises rather than to fundamental issues.
Spurious effects may show up when increasing the sensitivity, as for example the source masses' $\chi$ modulating the Earth magnetic field, and mimicking the signal. 
We stress that a good magnetic screening is a key part of this setup, and of an upgraded one, as it must have the lowest noise and larger shielding factor.

\section{Conclusions and outlook}
\label{sec:conclusion}
In conclusion, we have designed and operated an experimental setup to search for spin-mass long-range forces based on a SQUID-magnetometer with state-of-the art sensitivity.
Our search excludes the presence of pseudoscalar particles within a large region of the coupling-mass parameter space, which is of cosmological relevance for the physics case for axions and ALPs \cite{MARSH20161}. 

Thanks to this work, we believe that the combined advances related to an augmented and enhanced setup may lead to an unprecedented sensitivity level. Such setup will exclude or confirm the presence of pseudo-Goldstone bosons, like axions, predicted by current theories beyond the standard model, paving the way to a promising technique to search for the so-called ``invisibles'' particles in an extremely broad mass interval.
Different experiments are searching for ALPs \cite{PhysRevLett.122.121802,shaft} by exploiting the exceptional magnetic field sensitivity obtainable with SQUID-based magnetometers, and rapid progress can be foreseen in approaching problems like the study of magnetic noise \cite{doi:10.1063/1.1147514} or the optimisation of the readout chain \cite{doi:10.1063/1.4976823}.

Eventually, we suggest that this kind of apparatus could be used for other fundamental searches, and may considerably improve existing results. 
If Dark Matter is constituted by very low mass ALPs, a persistent pseudomagnetic field at the frequency of the ALP mass will drive the GSO electrons \cite{PhysRevD.88.035023}. This result already improves by a factor 2 existing limits \cite{PhysRevLett.103.261801} above 1\,mHz in the noise-free parts of the spectrum (see Fig.\,\ref{fig:new}), and can be further optimised by removing the electric motor and increase the measurement band.
The use of polarized sources \cite{speake2} may extend the experimental sensitivity to the spin-spin interaction. As it does not violate parity and time reversal, its coupling constant is much larger than that for spin and mass, making it preferred for axion detection \cite{PhysRevLett.113.161801}.
We finally mention that a closely similar, and possibly simpler, setup could be used to probe the gravitoelectromagnetic effects of Earth rotation on the elementary spin \cite{MASHHOON1974,Mashhoon_2000}, or the local presence of moving magnetic monopoles \cite{PhysRevLett.48.1378,Dusad2019}.

\section*{Acknowledgements}
The authors want to acknowledge Mario Tessaro for his work on the superconducting wiring, on the electronics of the setup, and on the engine stabilization. We also thank Fulvio Calaon and Enrico Berto for their help with the construction and test of the apparatus, and Ruggero Pengo for the advices on the cryogenics. The Cryogenic Service of the Laboratori Nazionali di Legnaro is acknowledged for providing us large quantities of liquid helium upon request. We deeply thank the AURIGA team for sharing their knowledge on mechanical and electromagnetic noise reduction.
We also thank Yevgeny Stadnik for providing us with a theoretical point of view on the implications of these interactions. Eventually, we deeply acknowledge the Laboratori Nazionali di Legnaro for hosting and encouraging the experiment.

\begin{appendix}

\section{Source and signal}
\label{app:a}
The source masses are full lead cylinders enclosed into thin aluminium capsules, which are attached to the wheel.
To put the masses in motion we use a high torque brushless motor phase-synchronized with a signal generator to precisely control the wheel rotational frequency.
The motor shaft and the wheel pivot are connected by two pulleys whose teeth number is such that the harmonics of the motor rotation does not fall on the signal frequency. The used gearing ratio is 22/48.
This scheme ensures that the wheel's rotational, and thus the signal's, frequency is somewhat distanced from the motor's own frequencies. Motor-related disturbances may be due to the specifics of the drive mechanism, and we verified that they are present in the broadband spectrum of the SQUID noise (see Fig.\,\ref{fig:new}) only if the motor is turned on. Even if relatively large, they are found not to influence the measurement noise.
The masses  are embedded in a low-density foam to avoid the effects of air-currents, and the whole wheel plus motor system was enclosed in a thin aluminium cage to reduce the air movement around the cryostat.
Before starting the data acquisition, we initiate the wheel rotation and wait for some hours for the pulleys and motor to warm up in order to improve the frequency stability of the source.
We monitor the signal frequency with a capacitive sensor coupled to the wheel, from which we record the pseudomagnetic field frequency for the whole run. 

The displacement of the source masses creates the field modulation by changing $r$, the source-detector distance. 
For $\lambda\ll r$ the potential in Eq.\,(\ref{eq:v}) is exponential. If $\lambda\gg r$ it becomes a $1/r^2$ potential, which produces a signal modulation independent on the mediator mass. This is similar to what is usually done in experiments testing gravity, which has a $1/r$ potential with an infinite wavelength and no exponential decay.
Our wheel and masses were designed to have a signal modulation of about a factor 2 for $\lambda\simeq10\,$cm.
without the need of a much larger apparatus. This choice makes it easier to screen the external disturbances like turbulences, vibrations and electromagnetic fields. A larger source would have been more space-consuming, had a larger inertia and an overall increased control complexity.

In the favoured case of the detection of a signal, this may be tested by lifting the cryostat to vary the source-detector distance, in order to estimate the range of the fifth force and hence the mass of the mediator. In our apparatus this modification is straightforward and does not require additional work on the setup.

Through a Fourier transform of the simulated signal we verified that the spectral leakage, i.\,e. the fraction of signal lost in the harmonics, is negligible for $\lambda>1\,\mathrm{cm}$, as our expected signal is much closer to a sine wave (no signal leakage) than to a square wave (significant signal leakage).
\\

\begin{figure}
\centering
\includegraphics[width=.5\textwidth]{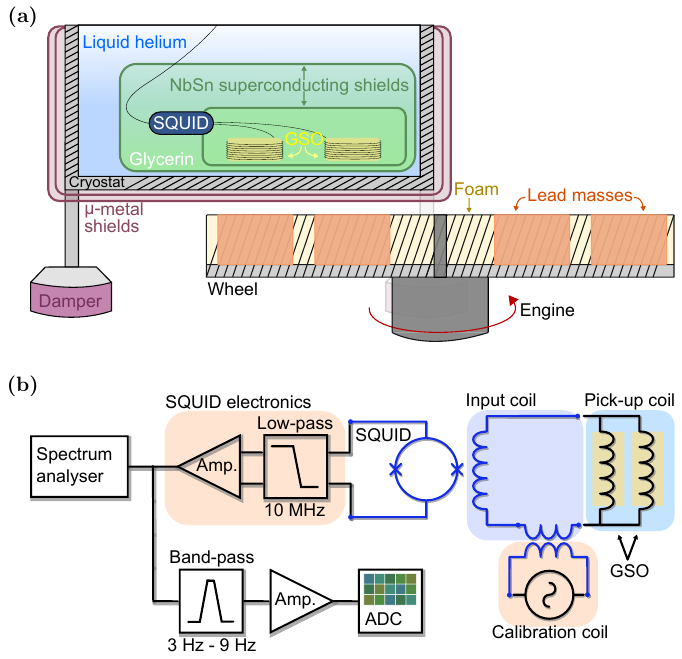}
\caption{Section of the apparatus and detection circuit. (a) Vertical section of the wheel, cryostat, and magnetic screens surrounding the detector. The drawing is not to scale. (b) Circuital scheme of the detector. The blue circuits are printed on a board, while the black ones are external wires and electronics. The input and pick-up coils are framed in different colours as they are considered two single condensed element in the main text, and longer coil symbols correspond to larger inductances. The GSO crystals are shown as yellow rectangles, Amp. are the amplifiers, and ADC is an analog-to-digital converter.}
\label{fig:5}
\end{figure}

\section{Detector and readout}
\label{app:b}
To match the $1.8\,\mu\mathrm{H}$ input impedance of the SQUID we connect in parallel two pick-up inductances of roughly $3.6\,\mu\mathrm{H}$.
The GSOs are housed in 3D-printed plastic supports, around which the pick-up coils are wound. Each inductance consists in 8 turns of NbTi wire distributed over the height of the crystal. To design the pick-up, we tested the coil plus support system at liquid helium temperature, and verify the minimum dimensions which guarantee an adequate thermal stress resistance.
The presence of the plastic supports increases the area of the coils with respect to that of the crystals'. We end up with a single coil surface of about $\Sigma\simeq11\,\mathrm{cm^2}$ and correct our limit accordingly. Within the uncertainties, the calculated values of inductances and impedances are in good agreement with the measured values.
The electronics of the SQUID system is reported in Fig.\,\ref{fig:5}, which is a complete version of the simplified scheme in Fig.\,\ref{fig:2}a. 

The SQUID output is filtered and amplified before being split between two channels. The first is used as a monitor using a signal analyser, while the second is further filtered and amplified before being acquired by an analog-to-digital converter.
The system is calibrated by the injection of a known current in a calibration coil (see Fig.\,\ref{fig:5}), whose contribution to the inductance of the input coil is small. Subsequently, the field sensitivity of the detector is obtained considering the geometric features of the pick-up coil.
The procedure's accuracy was tested in a previous work \cite{CRESCINI2017677} with an additional calibration, namely the use of a large solenoid to provide a controlled and uniform AC field over the pick-up coil. The two independent calibrations matched within experimental uncertainties, so for this work we relied on the reported one, and avoided the addition of other coils to the setup.

\section{Background estimation and signal analysis}
\label{app:c}
The SQUID noise at the pick-up depends on the inductances of the input coil, of the pick-up coil and on the mutual inductance between the input coil and the SQUID loop $M_i$.
The flux power spectral density of the intrinsic detector noise at $L_p$ can be calculated as \cite{squidhand}
\begin{equation}
S_\phi^\mathrm{(p)}(\omega) = \frac{(L_p+L_i)^2}{M_i^2}S_\phi(\omega).
\label{eq:squidnoise}
\end{equation}
The magnetic field noise can be calculated from Eq.\,(\ref{eq:squidnoise}) as $S_B^\mathrm{(p)}(\omega)^{1/2} =S_\phi^\mathrm{(p)}(\omega)^{1/2} / N_p n_p\Sigma$, where $N_p=2$ is the number of pick-up coils and $n_p=8$ is the number of turns of area $\Sigma$ forming each coil.
Our sensor has a flux noise per unit of bandwidth $S_\phi(\omega)^{1/2}\simeq 1\,\mu\phi_0/\mathrm{\sqrt{Hz}}$, input and pick-up inductances $L_i\simeq L_p \simeq 1.8\,\mu\mathrm{H}$, and SQUID-$L_i$ mutual inductance $M_i\simeq8.8\,\mathrm{nH}$, leading to an expected field noise  $S_B^\mathrm{(p)}(\omega)^{1/2} = 46\,\mathrm{aT/\sqrt{Hz}}$, which is in agreement with the measured background.
The energy resolution of our system may be calculated as
\begin{equation}
\epsilon = \frac{1}{2} L_i S_\phi(\omega)/M_i^2=443\,\hbar
\label{eq:enres}
\end{equation}
meaning that a quantum limited SQUID sensor, with $1\,\hbar$ energy resolution, could improve the present magnetic field sensitivity of a factor $\sqrt{\epsilon/\hbar}\simeq20$.

The magnetisation noise is expected to contribute to the noise budget of the experiment, and its magnitude can be quantified using the fluctuation-dissipation theorem as
\begin{equation}
S^\mathrm{(p)}_M(\omega)^{1/2} = \sqrt{\frac{2 k_B T}{\pi \mu_0}\frac{\chi \tau}{V}},
\end{equation}
where $V\simeq 14\,\mathrm{cm^3}$ is the crystals' volume, and $\tau \simeq 70\,\mathrm{ps}$ is the spin relaxation time of the GSO, measured at 80\,K through an electron paramagnetic resonance \cite{gso,Crescini2017109}.
With our experimental parameters the magnetisation noise results $12\,\mathrm{aT/\sqrt{Hz}}$, which is compatible with the added noise present in our spectra.
We tested the system with and without GSO crystals inserted. However, to satisfy the matching condition $L_p\simeq L_i$ in the two configurations, the number of turns of the coils have to be changed in order to compensate the higher inductance related to the crystals' magnetic susceptibility. This adds a systematic uncertainty in the comparison of the two measurements, which is difficult to take into account, and prevents us from reporting a differential result.
Referring to Fig.\,\ref{fig:2}d, we consider a frequency interval close to the pseudomagnetic signal frequency, to estimate the uncertainty of the measured noise level.
The noise related to the $\mu$-metal shielding is of the order of $10\,\mathrm{fT/\sqrt{Hz}}$ \cite{doi:10.1063/1.2885711} and, being a magnetic noise, it is screened by the superconducting shields of several orders of magnitude. Given the screening factor of the employed NbSn shields, we do not expect this noise to constitute a limit in this or in future upgrades of the apparatus.

In a complementary way, the noise spectra presented in Fig.\,\ref{fig:2}d and in Fig.\,\ref{fig:3} demonstrate the absence of a persistent signal in our data set. The former shows that with the minimum bandwidth allowed by the experimental integration time there is no trace of a signal, while the latter ensures the stability of the measurement throughout the run. The small field excess in Fig.\,\ref{fig:2}d at 6.05\,Hz is justified by the data in Fig.\,\ref{fig:3}b, which also displays that this increase is not persistent, is not at the pseudomagnetic field frequency, and is not contained within one frequency bin. We conclude that it can not be confused with a signal.

\section{Spurious noise reduction}
\label{app:d}
In this experimental scheme the working frequency can be precisely managed by varying $f_w$, which is a major advantage in the control and reduction of the external noises.
The nature of these disturbances is usually electromagnetic or mechanical; hereafter we list the different measures we took to reduce such effects.

Electromagnetic disturbances are related to the environment and to the instruments used in the experiment, like the motor driving the wheel.
As shown in Fig.\,\ref{fig:5}a, the design of our apparatus was optimised to get the minimum source-detector distance while preserving a satisfactory screening of the magnetometer. The details of the magnetic shields are as follows. 
Two $\mu$-metal shields are placed around the cryostat at room temperature, and consist in thin sheets of $\mu$-metal shaped in-situ and not re-baked or soldered.
Two superconducting shields are immersed in liquid helium, one comprises a large volume around the detector and the second only contains the crystals and pick-up coils. These cryogenic shields are made of single lead-tin sheets, with no soldered joints. 
The shielding factor of the same material batch which constitutes our shields was measured in a dedicated experiment, and for two layers the screening factor results of the order of $10^{-10}$.
A third shield is dedicated to the SQUID chip, and is made of niobium.

The $\mu$-metal response to a weak electromagnetic field was treated for both the detection and screening of exotic interactions \cite{Vorobyov1988146,PhysRevLett.82.2439,Gohil:2696276,PhysRevD.94.082005}. It was discussed whether such a faint force could orient the magnetic domains of a ferromagnet, and initial proposals were rejected on the basis that it can not, driving the field toward the use of paramagnets \cite{PhysRevLett.82.2439}.
In this work, and in particular in the consistency test described in Section \ref{sec:setup}, we found that the magnetic screening in our apparatus at 6\,Hz is essentially due to the superconducting shields. Since the $\mu$-metal does not seem to play a significant role, we argue that a weak magnetic field is not orienting the domain of the shield. We stress that this finding holds in our apparatus, whose shield consists in $\mu$-metal foils formed on-site, and might differ from results obtained with thicker shields like to ones typically used in comagnetometers \cite{PhysRevD.94.082005}. Furthermore, we argue that the cancellation of an exotic interaction due to $\mu$-metal is not straightforward in the presence of superconducting shields, and for fifth forces with a potential not proportional to $1/r$. This will be detailed in a future work.
In the present setup, the $\mu$-metal is useful to allow the superconducting transition of the screens to occur in a reduced static magnetic field, improving their rejection factor and reducing the trapped field.
Therefore, the presence of $\mu$-metal between the source and the detector can be avoided by enclosing both of them in a single magnetic shield, or working in a magnetically shielded room.

Considerable effort was made to eliminate possible electromagnetic cross-talk between the power electronics of the motor and the magnetometer, and for the same reason the square wave reference signal given by the capacitive sensor of the source is smoothed with a low-pass filter.
Some other techniques that we used to avoid the presence of spurious signals at the pseudomagnetic field frequency were presented in the description of the source system.

Mechanical vibrations are driven by the wheel rotation, and by the external environment, so, to reduce them, we isolate the cryostat from the source and from the surroundings.
The cryostat was decoupled from the ground with four legs resting on alternate layers of lead and Sylodamp\textsuperscript{\textregistered} SP1000, a polyurethane elastomer which damps vibrations.
Turbulences and acoustic disturbances related to the masses rotation are diminished thanks to the geometry of the source, which is made as uniform as possible, and by enclosing the motor plus wheel system in a thin aluminium cage.
A relative movement between the detector and the static field trapped in the superconducting shields is particularly problematic, as it leads to a magnetic flux in the pick-up coils.
To drastically reduce this issue we fill the chamber containing the screens and the coils with glycerin, which freezes at low temperature and blocks all the relative vibrations.

\begin{figure}
\centering
\includegraphics[width=.5\textwidth]{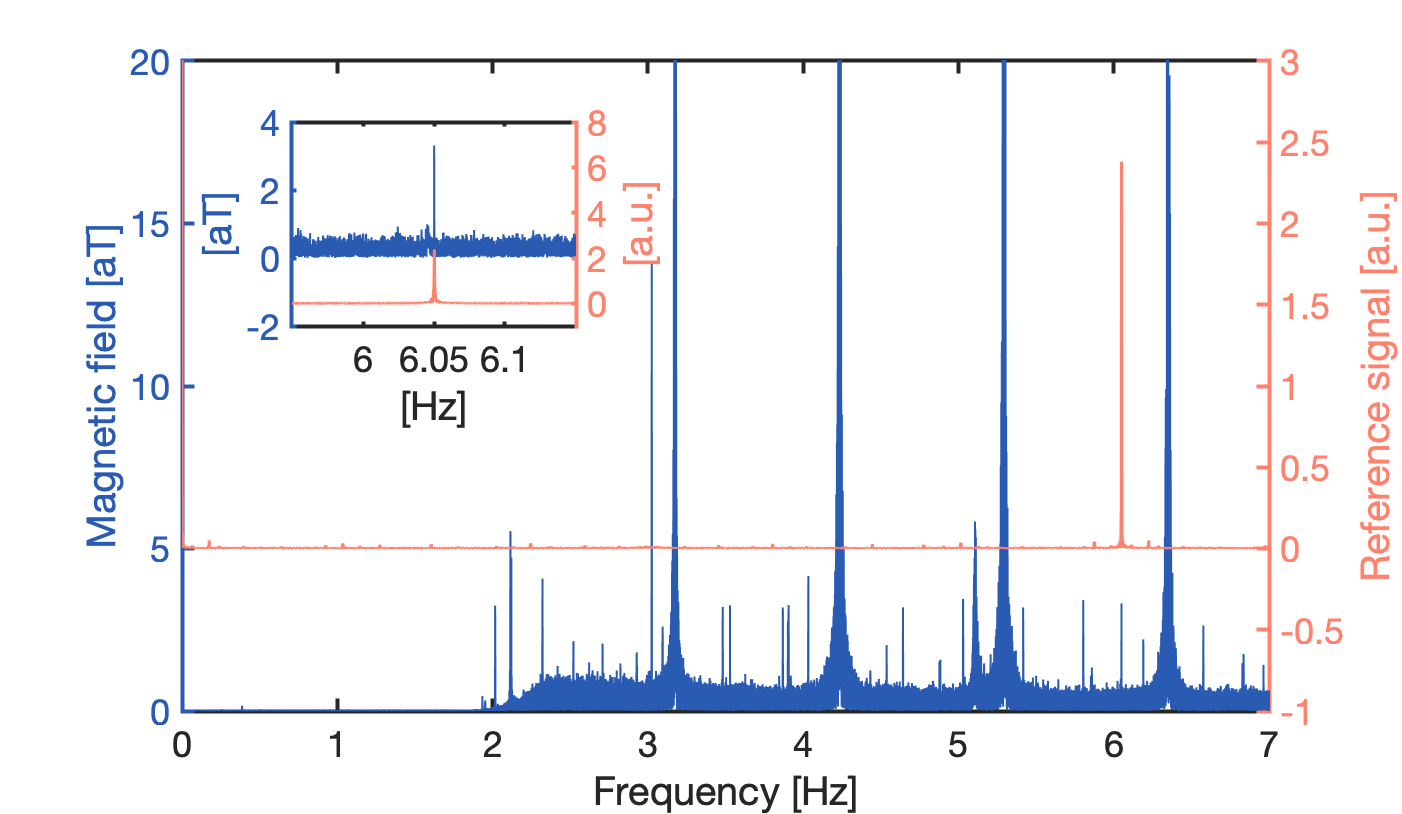}
\caption{Magnetic noise and signal spectrum of a run where an electromagnetic signal was injected on purpose in the apparatus. The actual frequency for the reference signal (orange curve) can be derived by dividing the $x$-axis values by $n_s=12$. The SQUID magnetic noise spectrum (blue) is affected by several noise peaks. Although the ones with higher amplitudes are known to be related to the electric engine, not all the other spurious peaks' origins were clearly identified. The signal is at 6.05\,Hz, in band free of magnetic disturbances, and shows no other harmonics down to 0\,Hz. The inset shows the same spectra in a reduced band. For further details, see the main text and Appendix \ref{app:b}.}
\label{fig:new}
\end{figure}

Moreover, the result of a run where this synthetic fifth force for added is reported in Fig.\,\ref{fig:new}, where the reference signal spectrum was overlapped with the magnetic one.
The former is not affected by significant sub-harmonics, and shows the bin where the signal is detected. In the latter, several noise peaks are present, and the low frequency cut-off is due to the bandpass filter. The main noise peaks are related to the engine, while the minor ones were not studied as their frequencies are much different than the one of the signal. Overall, none of these disturbances affects the signal bin.

\end{appendix}

\bibliographystyle{plain}
\bibliography{gpgsQUAX2020}

\end{document}